\documentstyle[12pt,a4]{article}
\begin{document}
\newcommand{\wst}{~{}^{{}^{*}}\llap{$\it w$}}
\newcommand{\wdst}{~~~{}^{{}^{**}}\llap{$\it w$}}
\newcommand{\omegast}{~{}^{{}^{*}}\llap{$\omega$}}
\newcommand{\omegadst}{~~~{}^{{}^{**}}\llap{$\omega$}}
\newcommand{\must}{~{}^{{}^{*}}\llap{$\mu$}}
\newcommand{\mudst}{~{}^{{}^{**}}\llap{$\mu$~}}
\newcommand{\nust}{~{}^{{}^{*}}\llap{$\nu$}}
\newcommand{\nudst}{~{}^{{}^{**}}\llap{$\nu$~}}
\newcommand{\beq}{\begin{equation}}
\newcommand{\eeq}{\end{equation}}
\newcommand{\gfrc}[2]{\mbox{$ {\textstyle{\frac{#1}{#2} }\displaystyle}$}} 
\title {\large \bf De-Sitter Type of Cosmological Model in n-Dimensional Space - Time - Mass Theory of Gravitation }
\author{{\bf G S Khadekar} \thanks{Tel.91-0712-23946,
email:gkhadekar@yahoo.com} and {\bf Vrishali Patki} \\
Department of Mathematics, Nagpur University \\
Mahatma Jyotiba Phule Educational Campus, Amravati Road  \\
Nagpur-440010 (INDIA) }
\maketitle
\begin{abstract}
Exact solution are obtained for  a homogeneous spapcially isotropic cosmological model in a matter free space with or without cosmological constant for a n-dimensional Kaluza-Klein type of metric in the  rest mass varying theory of gravity proposed by Wesson[1983]. The behavior of the model is discussed. 
\\
\\
{\bf Key Word}: Kakuza-Klein theory, Wesson theory, Higher dimensional space time.\\
{\bf AMS Subject Classification }: 83EXX 
\end{abstract}
\section{ Introduction}
 \par In a past few years there have been many attempts to construct a unified field theory based on the idea of   multidimensional space time. In-need it is generally be-lived that higher dimensional must play a significant role in the early universe.  Wesson [1983] proposed a 5-dimensional Space-Time-Mass (STM) theory of gravity  with variable rest mass. In this theory 5 coordinates are $ x^0 =ct,$ the three space co-ordinates are $ x^1 = x, \:  x^2 = y,\: x^3 = z$ and $ x^4=\frac{Gm}{C^2}$, this new 5-dimensional theory of variable rest  mass as a natural extension of general theory of relativity. The existence of constant $ C $ suggest that $ x^4 = \frac{Gm}{C^2} \; (m=mass)$ be a coordinate in 5-dimensional theory of space- time- mass. \\
 \par In our present work we extended the work of Chaterjee[1987] for n-dimensional variable mass theory of gravity. We have obtained an  exact solutions 
for a homogeneous spatially  isotropic n-dimensional cosmological model in vacuume both with or without cosmological constant. However, it is pointed out that Chaterjee's[1987] solution is a particular case of solution presneted here.      
\subsection{Field Equations}
 The line element for a n-dimensional  homogeneous and
spatially isotropic  cosmological model is taken as \\
\begin{equation}
ds^2 = e^\nu dt^2-e^{\omega} dx^2 + e^{\mu} dm^2
\end{equation}
where $ dx^2 =\sum_{i=1}^{(n-2)}dx^2_{i} $ and  $\mu,{\omega} $ and $\nu $ are the functions of time and mass. Here the coordinate $x^0 = t, \:
 x^{1,2, \cdots, (n-2) } $  (space  coordinate) and $x^{(n-1)} = m.$ For simplicity we have set the magnitudes of both $C$ and $ G$ to unity. By applying this metric to the Einstein field equation
 $ G_{ij} = R_{ij} -\gfrc{1}{2} g_{ij} \, R = \Lambda g_{ij} $ with the assumption
$e^{\nu}= 1$,  we get
\beq
G_{00} = -(n-2)(n-3) \; \frac{\dot \omega^2}{8}- (n-2) \: \frac{\dot \omega\dot 
\mu}{4} - (n-2) \, e^{-\mu}\left( \frac{\omegadst}{2}-\frac{\omegast \must}{4} + (n-1) \, \frac{\omegast^2 }{8}  \right)= -\Lambda
\eeq
\begin{eqnarray}
G_{11} = e^{\omega} \left(\frac{\ddot \mu}{2}  + \frac{\dot \mu^2}{4} 
\right) +  (n-3)e^{\omega}  \left(\frac{\ddot \omega }{2} +\frac{\dot{ 
\omega} \dot {\mu}}{4} +(n-2)\frac{{\dot \omega^2} }{8} \right)
 \nonumber \\ +  (n-3)  e^{\omega- \mu} \left( \frac{\omegadst }{2} -
 \frac{ \omegast \must}{4} +  (n-2)  \frac{ \omegast^2}{8} \right)= \Lambda  
\end{eqnarray}
 $$ G_{11} = G_{22} = G_{33} = \cdots = G_{(n-2)(n-2)} $$
\beq
G_{0(n-1)} = (n-2) \left( \frac{\dot{\omega}^*}{2 }+ 
\frac{\omegast \: \dot{ \omega}}{4}-\frac{\dot \mu \omegast}{4}\right) = 0 
\eeq
\beq
G_{(n-1)(n-1)} = -(n-2)(n-3) \frac{\omegast^2}{8} - (n-2) \;  
e^{\mu}\left(\frac{ \ddot \omega}{2} +(n-1)\frac{\dot{\omega^2}}{8}\right)=- 
\Lambda e^{\mu}
\eeq
 where  a dot$(.)$ and star$(*)$ denote, respectively partial derivative with respect to time and mass.
\subsection{Solutions}
By solving equation (4) we get
\beq
 e^{\mu}= \omegast^2 \frac{e^{\omega}}{\alpha(m)}
\eeq
where $\alpha(m)$ is an arbitrary function of mass only.\\ Since $\omegast \ne  0 $ we get,  using equation (6) in (5) 
\beq
e^{\omega}\left[{\ddot \omega} 
+ \frac{(n-1)}{4} \; {\dot \omega^2}-\frac{ 2\Lambda}{(n-2)}\right] + \frac{(n-3)}{4}\alpha(m) = 0
\eeq
Replacing $ e^{\omega}$  by  $ y$ we get
\beq   {\ddot y}- \frac{2\;\Lambda}{(n-2)}y + \frac{(n-3)}{4}\; \alpha = 0
\eeq
After first integration we get 
\beq
\dot{y^2}= \frac{8\; \Lambda}{(n-1)(n-2)}y^2 - \alpha \; y + 
\frac{\gamma}{y^{\frac{(n-5)}{2}}}
\eeq
where $\gamma$ is a arbitrary function of mass only.\\ Different cases for the equation (9) may arise. We shall consider four of them.\\
\\
{\bf Case I} : $ \Lambda = 0, \; \gamma = 0 $ \\
In this case simple solution of equation (9) is
\beq 
e^{\omega} = \alpha_{1}\;t^2 + \beta_{1}\; t +  \gamma
\eeq
where $ \alpha_{1} =\frac{-\alpha}{4}\; $ , $\beta_{1}$ and $ \gamma_{1} $ are arbitrary functions of mass only. This result is exactly identical to those obtained by Chaterjee [1987] for 5-dimension.\\
\\
{\bf Case II }: $ \Lambda = 0 ,\;  \gamma \ne 0$ \\
After an extremely tedious but straight forward calculation we get the general solution of equation (9) as
\beq
\frac{2}{\alpha^{N_{1}}}\left[\frac{- \;1}{(\gamma - x^2)^{\frac{N_{2}}{2}}       } + N_{2}\;  \gamma  \int{\frac{dx}{(\gamma - x^2)^{N_{3}}}} \right]= t + C_{1}
\eeq
where $ C_{1} $is an arbitrary function of mass and $ \;  x^2 = (\gamma - \alpha 
e^ {\frac{(n-2)\omega}{2}}),\; N_{1} = \frac{(n-1)}{2(n-3)}, \; N_{2} = \frac{(n-5)}{(n-3)}, \; N_{3}= \frac{(3n-1)}{2(n-3)}$ \\
\\
{\bf Case III} : $\Lambda \ne 0 \; \; \gamma = 0 $\\
In this case the solution of equation  (9) is 
\beq
\frac{4 \Lambda}{3}e^{\omega}- \alpha + 2\;(\frac{2 \; \Lambda}{3})^{\frac{1}{2}}
\left[ \frac{2 \; \Lambda e^{2\omega}}{3}- \alpha e^{\omega}\right]^{\frac{1}{2}} = e^{(\pm t +C_{2})}
\eeq
where $ C_{2} $ is an arbitrary function of mass only.\\ This solution is formerly the same as the the solution obtained by Chaterjee[1987] for 5-dimension.\\
\\
{\bf Case IV} : $ \Lambda \ne 0 \; \; \gamma \ne  0 $\\
From equation (9) we get the general solution
\beq
 \frac{e^{K_{1} \; \omega}}{K_{1}} - \frac{ K_{2} \; e^{K_{3}\; \omega}}
{ 2 \;\gamma \; K_{3}} + \frac{ \alpha  \; e^{K_{4}\; \omega}}
{ 2 \;\gamma \; K_{4}} = (\gamma)^{\frac{1}{2}} \; t + C_{3}
\eeq
where $ K_{1} = \frac{(n-1)}{4} , \; K_{2} = \frac{8\;  \Lambda}{(n-1)(n-2)},\; K_{3}= \frac{3 \;(n+1)}{4}, \; K_{4} = \frac{(3n-7)}{4}$
\subsection{Conclusion}
\par In this paper we have considered the the n-dimensional Kaluza-Klein type metric in rest mass varying theory of gravity. The solution obtained here is more general by Chterjee[1987] for 5-dimensional case. Chaterjee's solution is a particular case of the solution presented here . We think that this new exact higher dimensional solution together with cosmological consideration should bring some additional information and as such they need to be further investigated. It is our hope that the higher dimensional solution presented here can be used as the starting point to investigate the behavior of the rest of the particles in more realistic universe model. 
\bibliographystyle{plain}

\end{document}